\documentclass[aps,pra,twocolumn,showpacs,amsmath,amssymb]{revtex4}
\usepackage{amsmath}
\usepackage{amsfonts}
\usepackage{float}
\usepackage[final]{epsfig}

\newcommand{\BesselJ}{\operatorname{J}}
\newcommand{\Ai}{\operatorname{Ai}}
\newcommand{\rmd}{{\rm d}}
\newcommand{\rme}{{\rm e}}
\newcommand{\rmi}{{\rm i}}

\begin{document}
\title{Observed photodetachment in parallel electric and magnetic fields}
\author{John N.~Yukich}
\email{joyukich@davidson.edu}
\homepage{http://webphysics.davidson.edu/faculty/jny/welcome.htm}
\affiliation{Physics Department, PO Box 7133, Davidson College,
Davidson, NC 28035-7133}
\author{Tobias Kramer}
\email{tkramer@ph.tum.de} \affiliation{Physik--Department T30c,
Technische Universit\"at M\"unchen, James--Franck--Str., 85747
Garching, Germany}
\author{Christian Bracher}
\affiliation{Department of Physics and Atmospheric Science, Dalhousie
University, Halifax, N.S.\ B3H~3J5, Canada}

\date{\today}

\begin{abstract}
We investigate photodetachment from negative ions in a homogeneous
1.0 Tesla magnetic field and a parallel ac electric field of $\sim$
10 V/cm. A theoretical model for detachment in combined fields is
presented. Calculations show that a field of 10 V/cm or more should
considerably diminish the Landau structure in the detachment cross
section. The ions are produced and stored in a Penning ion trap and
illuminated by a single-mode dye laser. We present preliminary
results for detachment from $\rm S^{-}$ showing qualitative agreement
with the model.  Future directions of the work are also discussed.
\end{abstract}


\maketitle

\section{INTRODUCTION AND MOTIVATION}
\label{sec:intro}

One of the long-term objectives of photodetachment studies has been
to shed light on electron correlation effects and other electron-atom
interactions.  One way to enhance photodetachment studies is to add
external electric or magnetic fields.  These potentials create
conditions in negative ions not found in neutral atoms, and permit
comparisons with phenomena observed in neutral atom photoionizaton.
For example, photodetachment in a static electric field environment
releases an electron that experiences an acceleration analogous to
that of freefall.  External fields permit studies of interference
phenomena and provide insight into the connection between quantum
mechanics and semiclassical closed-orbit theories.  A large amount of
attention, both theoretical and experimental, has been given to
photodetachment in the presence of external fields
\cite{manakov2000,blumberg1,blumberg2,fabrikant1981,demkov1982,
bryant88,baruch90,baruch92,gibson93a,gibson93b,gibson2001,delos88a,
delos88b,fabrikant94,greene88,wong88,krause,bryant87,fabrikant91,
starace97,delos93a,delos93b,delos94,delos97,wang97a,wang97b,kramer2001,
blondel96,blondel99,kramer2002,bracher2003}.

Photodetachment can be thought of as the latter half of a collision
between an electron and a neutral atom.  In general, the effect of an
external field is to cause a portion of the out-going electron wave
function to return to the atomic core, where additional interactions
may occur.  For example, an electric field creates a potential slope,
and the electron wave function travelling uphill is eventually
reflected and revisits the core.  If the coherence time of the
transition is long compared to the reflection time, interference
occurs between the parts of the wave function emitted in the upward
and downward directions. The time difference, and thus the relative
phase between the two waves, depends on the detached electron's
energy and momentum, and diminishes with increasing applied field.
The interference gives rise to oscillatory structure in the
detachment cross section depending on the relative phase between the
waves.  The effect has been observed and discussed in numerous papers
\cite{fabrikant1981,demkov1982,bryant88,baruch90,baruch92,gibson93a,
gibson93b,gibson2001,blondel96,blondel99,kramer2002,bracher2003}.
Throughout the reflection process the electron wave function spreads,
unconstrained, in the directions perpendicular to the field. Thus,
for a given energy above threshold, a larger electric field causes a
larger portion of the wave function to revisit the core, yielding a
larger interference amplitude. Most of the experimental observations
have employed a field on the order of $\rm 100~V/cm$ or more in order
to resolve the oscillatory structure in the detachment rate
\cite{bryant88,baruch90,baruch92,gibson93a,gibson93b,gibson2001}.

A similar interference occurs in the presence of a magnetic field,
whose effect is to constrain and quantize the motion of the outgoing
electron in the plane normal to the magnetic field \cite{Landau}. The
field-free detachment threshold is replaced by a series of Landau
thresholds in the cross section at uniformly spaced photon energies
corresponding to the cyclotron states \cite{blumberg1,blumberg2}.
This periodic structure in the cross section arises from the outgoing
electron wave function constructively interfering with itself, as the
orbiting electron revisits the core once every cyclotron period.
This interference structure is in one sense more apparent than that
produced by an electric field because the magnetic field confines the
electron in two dimensions, and consequently a larger portion of the
wave function returns to the core.  Similar structure has also been
observed in photoionization in a magnetic field
\cite{Reinhardt83,delos87,delos88a}.

Interference is suppressed if the cyclotron-orbiting electron does
not strictly return to the core.  This situation can be brought about
in a couple of ways.  One is to introduce a small, static electric
field parallel to the magnetic field.  Such a field tends to
accelerate the electron away from the ion core along the fields'
axes, diminishing or even eliminating the dramatic interference of
the cyclotron states \cite{delos94,delos97,wang97a,wang97b}. The
electric field effectively ``opens'' the closed cyclotron orbits.
Earlier quantum mechanical calculations \cite{du89,fabrikant91}, as
well as those in this paper, show that a field on the order of 10
V/cm should significantly reduce the Landau structure observed in the
detachment cross section.  A commensurate prediction can be made from
a semiclassical calculation assuming that, in order to have
constructive interference, the electron must return to the atomic
core within half a Bohr wavelength in one cyclotron period.
Furthermore, the motional Stark field experienced by a thermally
energetic ion affects the interference.  In the ion's rest frame, a
motional electric field perpendicular to the magnetic field causes
the electron to drift away from the core, diminishing resolution of
magnetic field structure in the photodetachment rate.  Again,
theoretical calculations, both quantum mechanical and semiclassical,
predict that a motional field on the order of 10~V/cm or more tends
to wash out the cyclotron structure found in the magnetic field-only
cross section \cite{fabrikant91,delos93a}.

Experimentally, it is common for stray electric fields to be present
(such as that used in a Penning trap, discussed below, or motional
Stark fields), so it is important to understand how such fields
affect the detachment spectrum.  Although detachment in combined
external fields has been discussed extensively in the literature, it
has received little experimental attention
\cite{bryant87,fabrikant91,starace95,
starace97,delos93a,delos93b,delos94,delos97,du89,wang97a,wang97b}. In
this paper we present a theoretical prediction for detachment in
combined, parallel external fields. We then describe an experiment to
observe detachment from $\rm S^{-}$ in parallel fields of 1.0~Tesla
and 10~V/cm. Preliminary results are shown to be in at least
qualitative agreement with the model, and finally, future directions
for this work are discussed.

\section{Theoretical Description}
\label{sec:theory}

For a theoretical analysis of the photodetachment rate, we employ the
quantum source formalism.  In this approach the scattering event is
divided into two separate stages, absorption of a photon and
subsequent emission of the photoelectron into the external field
environment.  Details of the method, which is equivalent to the leading
order of conventional scattering theory, are presented in a recent
publication \cite{bracher2003}.  In near-threshold photodetachment,
the emitting ion is conveniently described as a pointlike source of
electrons, the dynamics of which follow from the quantum propagator
for the external potential.  Particularly simple expressions are
found for s-wave detachment in a uniform electric field
\cite{kramer2002}.  There, our predictions for both the total
detachment cross section and the spatial distribution of the
photoelectrons are in excellent agreement with the photocurrent
spectrum recorded by Gibson et~al.\ \cite{gibson93a,gibson2001} and
the photodetachment microscope images recorded by Blondel and
colleagues \cite{blondel96,blondel99}, respectively.

\subsection{Source model of photodetachment}
\label{sec:theory1}

In the quantum source picture, the motion of a photoelectron that is
emitted with an initial kinetic energy $E$ close to zero is governed
by the modified Schr\"odinger equation:
\begin{equation}
\label{eq:th1} \left\{ E - H(\mathbf r,\mathbf p) \right\}
\psi(\mathbf r) = C\cdot \delta (\mathbf r) \;.
\end{equation}
Here, we modelled the emitting ion as a pointlike entity, and
subsumed the details of the laser-ion interaction in the source
strength parameter $C$. (In $s$--wave detachment, the convenient
point-like form of the r.h.s.\ in Eq.~(\ref{eq:th1}) emerges from
the dipole interaction term $\boldsymbol\epsilon \cdot \mathbf D
\psi_{\mathrm{ion}}(\mathbf r)$ in the limit $E\rightarrow 0$, as
the extended initial wavelength of the photoelectron effectively
obliterates the detailed source structure.  See also
Ref.~\cite{bracher2003}.)  The solution to (\ref{eq:th1}) is given by
the energy Green function $G(\mathbf r,\mathbf o;E)$ describing an
outgoing wave in the external potential: $\psi(\mathbf r) = C\cdot
G(\mathbf r,\mathbf o;E)$.  The energy Green function, in turn, is
linked to the familiar time-dependent quantum propagator $K(\mathbf
r,t|\mathbf o,0)$ by a Laplace transform \cite{Economou}:
\begin{equation}
\label{eq:th2} G(\mathbf r,\mathbf o;E) = -\frac{\rmi}{\hbar}
\int_0^\infty \rmd t\; {\rm e}^{{\rm i}Et/\hbar} K(\mathbf
r,t|\mathbf o,0) \;.
\end{equation}
In this context, we are merely interested in the total particle
current transported by the wave $\psi(\mathbf r)$.  Integration over
a surface enclosing the point source immediately yields an expression
for the detachment rate:
\begin{equation}
\label{eq:th3} J(E) = -\frac2\hbar |C|^2 \lim_{\mathbf r\rightarrow
\mathbf o} \Im\bigl[ G(\mathbf r,\mathbf o;E) \bigr] \;.
\end{equation}
A general expression for the quantum propagator of electrons in
uniform electric and magnetic fields at arbitrary angles was given by
Nieto \cite{Nieto92}.  After insertion in (\ref{eq:th3}), we find
from (\ref{eq:th2}) the detachment rate:
\begin{widetext}
\begin{eqnarray}
J(E) & = & - \frac{2\omega_L}{\hbar^2} \left( \frac m{2\pi\hbar}
\right)^{3/2} |C|^2 \,\, \Im \nonumber \\
& & \times \left[ \int_0^\infty \frac{\rmd t}{\sqrt{{\rm i}t}
\sin(\omega_L t)} \exp\left\{ \frac{\rmi}{\hbar}\left( Et -\frac{e^2
F_{\parallel}^2 t^3}{24 m} +\frac{e^2 F_{\perp}^2 t}{8m\omega_L^2}
\bigl[ \omega_L t\cot\left(\omega_{L}t\right) -1 \bigr]
\right)\right\} \right] \;, \label{eq:th4}
\end{eqnarray}
\end{widetext}
where $\omega_L=eB/(2m)$ is the Larmor frequency, and
$F_\parallel,F_\perp$ denote the electric field components parallel
and perpendicular to the magnetic field, respectively.

While Eq.~(\ref{eq:th4}) yields the detachment rate for a stationary
ion undergoing any particular transition in an arbitrary field
geometry, we have to keep in mind that in the actual experiment, (i)
the ions are moving in the trap, (ii) an external electric field
accelerates the ions, and, (iii) transitions between
different magnetic sublevels with generally differing source strength
parameters $C$ and electron excess energies $E$ are occurring
simultaneously.  We assess these complications below.

\subsection{Thermal motion of the ions}
\label{sec:theory2}

The source model, formally expressed in the stationary Schr\"odinger
Eq.~(\ref{eq:th1}), implicitly requires that the emitting ion is at
rest. Obviously, this condition is not met by the trapped ions which
possess an average kinetic energy $\langle E_{\rm kin} \rangle =
\frac32 k_BT$, where $T$ denotes the effective temperature of the ion
cloud.  To be specific, in the following we assume thermal
equilibrium, i.~e., the (kinetic) momentum distribution $P(\mathbf
p)$ of the cloud is given by Maxwell's familiar expression:
\begin{equation}
\label{eq:th5} P(\mathbf p)=\frac{1}{{(2\pi M k_B
T)}^{3/2}}\exp\left(-p^2/(2 M k_B T)\right) \;.
\end{equation}
Here, $M \approx 32\;$amu is the mass of the sulfur ions.  To obtain the
proper detachment rates for a moving ion, we switch to its rest frame
of reference.  The transformation between the laboratory frame and
the rest frame involves a Lorentz boost operation that alters the
electric and magnetic fields perceived by the ion \cite{Jackson}.
Since the ion momentum is small, $\beta = p/(Mc) \ll 1$, we neglect
quadratic and higher contributions in $\beta$.  This approximation
leaves the magnetic field $\mathbf B$ unchanged, while the electric
field $\mathbf F$ is augmented by a motional electric field that
is perpendicular to $\mathbf p$ and $\mathbf B$:
\begin{equation}
\label{eq:th6} \mathbf F(\mathbf p) = \mathbf F + \frac1M
\left(\mathbf p \times \mathbf B\right) \;.
\end{equation}
Clearly, the additional electric field compensates for the magnetic
Lorentz force exerted on the moving ion in the laboratory frame.

Beside the uniform fields present in the trap, the change of
reference frame also affects the frequency of the detaching laser
beam, and hence the photon energy $\hbar\Omega$.  The parameter $E$
appearing in the Green function $G(\mathbf r,\mathbf o;E)$
(\ref{eq:th2}) is the excess energy of the photoelectron, i.~e., the
difference between $\hbar\Omega$ and the electron affinity $E_A$ of
the negative ion.  The Doppler corrected electronic energy
$\hbar\Omega-E_A$ in the rest frame of the ion is given in leading
order by:
\begin{equation}
\label{eq:th7} E(\mathbf p) =
\left[\hbar\Omega-E_A\right]-\frac{\hbar\Omega}{Mc}\left(\hat{\mathbf
n}\cdot\mathbf p \right)\;,
\end{equation}
where $\hat{\mathbf n}$ denotes the unit vector in direction of the
laser beam.  The corrections (\ref{eq:th6}) and (\ref{eq:th7}) cause
a dependence of the detachment rate on the ionic momentum $\mathbf
p$: $J(\mathbf p) = J(E(\mathbf p),\mathbf F_\perp(\mathbf
p),F_\parallel,\mathbf B)$.  In the experiment, however, only the
average detachment rate $J_{\rm average}(\Omega)$ is accessible which
follows from $J(E)$ (\ref{eq:th4}) after integration over the thermal
Maxwell distribution (\ref{eq:th5}):
\begin{equation}
\label{eq:th8} J_{{\rm average}}(\Omega) = \int \rmd^3\mathbf p\;
P(\mathbf p) J(\mathbf p) \;.
\end{equation}
We finally mention that the detachment rate in principle depends not
only on the momentum $p$ of the ions, but also on their position
$\mathbf r$ in the trap.  After all, an electric quadrupole is used
to confine the ions in the axial direction which gives rise to an
additional electric field that varies linearly with $\mathbf r$.
However, this field is comparatively weak, and while it can be
incorporated into the source formalism without fundamental
difficulties, for the sake of simplicity we neglect it here.
Furthermore, this approximation facilitates direct comparison of the
source model with the theory put forward by Blumberg {\em et~al}.\
\cite{blumberg2}.

\subsection{Time-dependent external field}
\label{sec:theory3}

In section~\ref{sec:theory1}, we obtained an expression for the
detachment rate $J(E)$ (\ref{eq:th4}) that was derived under the
condition of a static electric field $\mathbf F$.  Obviously, this
field accelerates the ion cloud, and under truly static conditions,
its center would settle into a new equilibrium position  in the
quadrupole potential that effectively eliminates the electric field.
Our setup avoids this problem by driving the ions using an
oscillating external field whose frequency $\omega_{\rm ext}$
considerably exceeds the resonance frequency of the trap $\omega_{\rm
trap}$, but still falls far short of the typical frequencies of
electronic motion ($\omega_L \sim 10^{11}$~Hz) \cite{Footnote1}.
Under these circumstances, photodetachment takes place in a
quasi-static external electric field $\mathbf F_{\rm ext}(\alpha) =
\mathbf F_{\rm ext}\cos\alpha$, where $\alpha=\omega_{\rm ext}t$,
while the measurement extends over many periods of the electric field
amplitude.  Similar to (\ref{eq:th8}), the effective photodetachment
rate $J_{\rm eff}(\Omega)$ then is given by the average of $J(E)$
over a complete period of the phase angle $\alpha$ of the external
field:
\begin{equation}
\label{eq:th9} J_{{\rm eff}}(\Omega) = \frac1{2\pi} \int_0^{2\pi}
\rmd\alpha \; J_{{\rm average}}(\Omega,\alpha) \;.
\end{equation}
At this point, we note that the momentum distribution $P(\mathbf p)$
(\ref{eq:th5}) in the driving external field becomes dependent on the
phase angle $\alpha$: $P(\mathbf p,\alpha) = P(\mathbf p - e\mathbf
F_{\rm ext}\sin\alpha/\omega_{\rm ext})$.  However, in the
configuration used in our experiment, $\mathbf F_{\rm ext}$ is
oriented parallel to the magnetic field axis, while the laser beam
points in the perpendicular direction.  It immediately follows that
the modified momentum distribution does not affect the corrections
(\ref{eq:th6}) and (\ref{eq:th7}).  We thus may combine
(\ref{eq:th8}) and (\ref{eq:th9}) to find the effective detachment
rate for any individual transition:
\begin{equation}
\label{eq:th10} J_{{\rm eff}}(\Omega) = \frac1{2\pi} \int_0^{2\pi}
\rmd\alpha \int \rmd^3\mathbf p\; P(\mathbf p) J\left(E(\mathbf
p),\mathbf F_\perp, F_\parallel, \mathbf B \right) \;,
\end{equation}
where $\mathbf F_\perp = \frac1M \left(\mathbf p \times \mathbf
B\right)$ and $F_\parallel = F_{\rm ext}\cos\alpha$. Inserting the
integral representation (\ref{eq:th4}) for the intrinsic current
$J(E)$, we find that both integrations in (\ref{eq:th10}) can be
carried out analytically, as the integrand is Gaussian in the
momentum, and the phase average can be expressed as a Bessel function
of order zero \cite{Abramowitz}.  We obtain the lengthy result:
\begin{equation}
\label{eq:th11} J_{{\rm eff}}(\Omega) = |C|^2 \cdot \Im\left[
\int_0^\infty \rmd t\; f(t)\,{\rm e}^{{\rm i}g(t)/\hbar} \right] \;,
\end{equation}
where the functions $f(t)$, $g(t)$ read:
\begin{eqnarray}
\label{eq:th12} f(t) & = & - \frac{2\omega_L}{\hbar^2} \left( \frac
m{2\pi\hbar} \right)^{3/2} \nonumber \\
& & \times \frac{{\rm i}\hbar M}{\sqrt{{\rm i}t}\,\sin(\omega_Lt)
\bigl[{\rm i}\hbar M - mk_BT t \bigl(1-\omega_L t \cot(\omega_L
t)\bigr)\bigr]} \nonumber \\ & & \times \BesselJ_0 \left( \frac{e^2
F_{\rm ext}^2 t^3}{48 m\hbar} \right) \;,
\end{eqnarray}
and
\begin{eqnarray}
\label{eq:th13} g(t) & = & (\hbar\Omega - E_A)t - \frac{e^2 F_{\rm
ext}^2}{48 m}\,t^3 \nonumber \\
& & - \frac{\hbar^2\Omega^2 k_BT t^2}{2c^2\bigl[{\rm i}\hbar M -
mk_BT t \bigl(1-\omega_L t \cot(\omega_L t)\bigr)\bigr]} \;.
\end{eqnarray}
At zero temperature and vanishing electric field, these expressions
reduce to the original expression for the detachment rate
(\ref{eq:th4}).  At this stage, only the temporal integration
remains.  Choosing a suitable path of integration allows for accurate
evaluation of (\ref{eq:th11}).  We deform the integration path into
the complex plane (carefully avoiding the singularities of the
integrand) and perform the integration numerically.

\begin{figure}[t]
\begin{center}
\includegraphics[width=0.9\columnwidth]{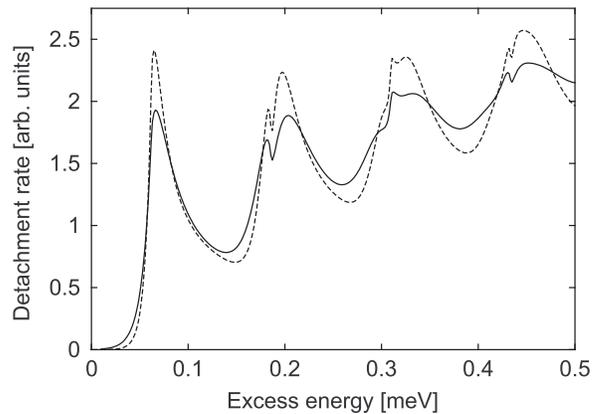}
\end{center}
\caption{Variation of the photodetachment rate $J_{{\rm
eff}}(\Omega)$ (\ref{eq:th11}) in a purely magnetic field as a
function of the excess energy $\hbar\Omega-E_A$ for various
temperatures.  Dotted line: $T=400$~K, solid line $T=950$~K (compare
with Ref.~\cite{blumberg2}, Figure~2). The broadening of the Landau
levels and some additional substructure due to the motional electric
field is visible. (Parameters used: Magnetic field: $B=1.07$~T, ion
mass: $M=32$~amu.)} \label{fig:Baverage}
\end{figure}
In Fig.~\ref{fig:Baverage} we show typical results for different
ionic temperatures in a purely magnetic field.  To compare with the
earlier results by Blumberg {\em et~al}., we used the same parameter set as
in Fig.~2 of Ref.~\cite{blumberg2} for the dashed curve.  Both
calculations are in excellent agreement, although they are obtained
by quite different methods:  Blumberg {\em et~al}.\ use a sum over Landau
levels to represent the current, and subsequently perform a numerical
thermodynamic average.  The more general approach presented here
allows to incorporate an electric, even time-dependent field at an
arbitrary angle.  Despite the inclusion of these additional features,
a single numerical integration suffices to accomplish the calculation
of the properly weighted detachment rate.  With decreasing temperature,
the Landau levels become more pronounced, while the additional
modulations visible in Fig.~\ref{fig:Baverage} persist. They may be
traced to the perpendicular motional electric field (\ref{eq:th6}).

\begin{figure}
\begin{center}
\includegraphics[width=0.9\columnwidth]{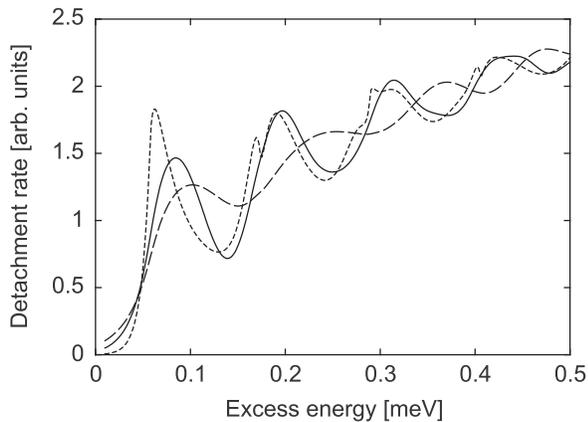}
\end{center}
\caption{Variation of the photodetachment rate $J_{{\rm
eff}}(\Omega)$ (\ref{eq:th11}) in parallel fields as a function of
the excess energy $\hbar\Omega-E_A$ for the electric field strengths
$F_{\rm ext} = 14$~V/cm (solid line) and $F_{\rm ext} = 28$~V/cm
(dashed line).  The dotted line represents the result for a purely
magnetic field.  (Magnetic field: $B=1$~T, ion mass: $M=32$~amu,
temperature: $T=950$~K.)} \label{fig:Fparallel}
\end{figure}
The effects of the additional oscillating parallel field $F_{\rm
ext}\cos(\omega_{\rm ext}t)$ are illustrated in
Fig.~\ref{fig:Fparallel}. Two major trends are easily discernible:
First, the resolution into discrete Landau levels loses contrast with
increasing electric field. In particular, the substructure seen in
Fig.~\ref{fig:Baverage} is absent.  For the higher value of the
electric field displayed in the figure ($F_{\rm ext} = 28$~V/cm), the
shape of the detachment rate spectrum is already reminiscent of the
``staircase'' appearance characteristic for $s$--wave detachment in a
purely electric external field
\cite{gibson93a,gibson2001,fabrikant94,bracher2003, kramer2002}.  The
other conspicuous feature in Fig.~\ref{fig:Fparallel} is a consistent
shift of the photocurrent maxima towards higher energies.  For a
qualitative explanation, we note that in a purely magnetic field, the
current $J(E,F_\parallel=0)$ (\ref{eq:th4}) generated by an $s$--wave
point source becomes singular at the Landau level thresholds $E_\nu =
(2\nu+1)\hbar\omega_L$:  $J(E,F_\parallel=0) \sim D(E-E_\nu)^{-1/2}$
\cite{blumberg2}, where $D$ denotes a constant.  (Inclusion of the
final-state interaction with the emerging neutral atom corrects the
unphysical behavior of the cross section at $E=E_\nu$, as discussed
in Refs.~\cite{clark83,sadeghpour00}.
However, the modification is of little practical relevance here
\cite{greene87}.)  An additional parallel electric field removes
the singularity and replaces the root by the square of an Airy
function \cite{fabrikant91,du89,starace95,kramer2001}:
\begin{equation}
\label{eq:th13a}
J_\nu(E,F_\parallel) = D \frac{\pi\sqrt{2m}}{\hbar\beta F_\parallel}
\Ai\bigl[ -2\beta(E-E_\nu) \bigr]^2 \;,
\end{equation}
where we introduced the parameter $\beta = \bigl[m/(2\hbar
eF_\parallel)^2\bigr]^{1/3}$.  As $\Ai(u)$ adopts its maximum at
$u_0=-1.018\ldots$ \cite{Abramowitz}, the maxima of the current
$J(E)$ are moved to energies $E = E_\nu - u_0/(2\beta)$.  We thus
expect a shift of the maxima in the averaged detachment rate $J_{{\rm
eff}}(\Omega)$ (\ref{eq:th11}) proportional to $\beta^{-1}$, i.~e.,
that scales with $F_\parallel^{2/3}$.  This conjecture is supported
by numerical calculations.

\begin{figure}
\begin{center}
\includegraphics[width=0.9\columnwidth]{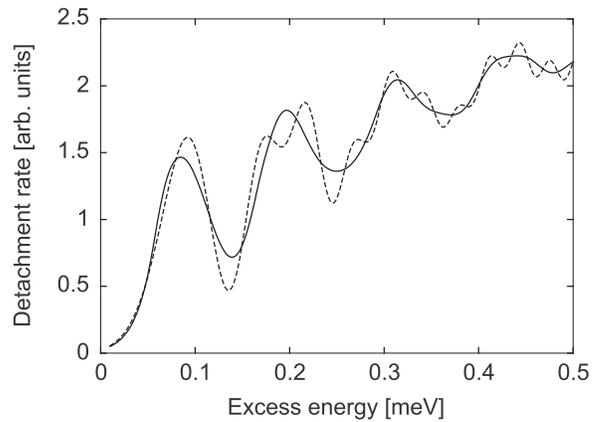}
\end{center}
\caption{Averaging effects caused by the oscillation of the parallel
electric field.  The figure shows the photocurrent spectrum $J_{{\rm
eff}}(\Omega)$ (\ref{eq:th11}) for a slowly oscillating field of
strength $F_{\rm ext} = 14$~V/cm (solid line) as a function of the
excess energy $\hbar\Omega-E_A$, as compared to the detachment rate
$J_{{\rm average}}(\Omega)$ (\ref{eq:th8}) in a constant external
field $F_\parallel = 10$~V/cm (dotted line). (Magnetic field:
$B=1$~T, ion mass: $M=32$~amu, temperature: $T=950$~K.)}
\label{fig:Oscillate}
\end{figure}

Much of the smooth appearance of the detachment rate in
Fig.~\ref{fig:Fparallel} is actually due to the averaging effect of a
time-dependent electric field, as Fig.~\ref{fig:Oscillate}
demonstrates.  The integration over the phase angle $\alpha$ of the
external electric field $F_{\rm ext}\cos\alpha$ (\ref{eq:th9})
effectively obliterates the oscillations in the photocurrent caused
by self-interference of the emitted electron wave
(Section~\ref{sec:intro}).  In fact, a comparison of the effective
rate $J_{{\rm eff}}(\Omega)$ (\ref{eq:th9}) for the parallel field
$F_{\rm ext}=14$~V/cm with the detachment cross section $J_{{\rm
average}}(\Omega)$ (\ref{eq:th8}) in a constant parallel field with
equal rms value ($F_\parallel = 10$~V/cm) shows coinciding overall
structure, while the numerous small-scale variations seen in the
constant field curve are missing in the averaged rate.

\subsection{Influence of magnetic substructure}
\label{sec:theory4}

Unfortunately it is not possible to measure the (averaged) total
current depicted in Figs.~\ref{fig:Baverage}--\ref{fig:Oscillate}
directly.  The magnetic field causes Zeeman
splitting of the magnetic sublevels of the ion and atom, and
consequently several transitions closely spaced in energy contribute
simultaneously to the photodetachment spectrum.  Here, we analyze the
situation encountered for the sulfur ions used in the experiment,
where the transition between the fine structure components
$^2P_{3/2}$ of the $S^-$ ion and $^3P_2$ of the emerging neutral $S$
atom was observed near the threshold at $E_A=2.0771~$eV
\cite{blumberg1,blumberg2}.

Denoting the initial states of the ion by $|J,L,S,M\rangle =
|\frac32,1,\frac12,M\rangle$, and the final states of the atom and
the detached electron by $|j,l,s,m\rangle=|2,1,1,m\rangle$ and
$|\frac12,m_s\rangle$ respectively, the relative frequency shifts
$\Delta\Omega(M,m,m_s)$ of the individual thresholds as compared to
the field-free case:
\begin{equation}
\label{eq:th14} \Delta\Omega(M,m,m_s) = (MG-mg-m_s g_s)\omega_L \;,
\end{equation}
depend on the Lande $g$--factor of the atomic entities which e.~g.\
reads for the ion:
\begin{equation}
\label{eq:th15} G = 1 + \frac{g_s-1}{2J(J+1))}\,
\bigl[J(J+1)+S(S+1)-L(L+1)\bigr] \;,
\end{equation}
where $g_s = 2.002319\ldots$ is the $g$--factor of a free electron.
The relative weight $|C(M,m,m_s)|^2$ (\ref{eq:th3}) of the various
transitions is proportional to the dipole matrix element between the
initial and final states; its calculation is a considerably more
complicated task that involves addition of the  angular momenta of
the ion, atom, photon, and electron \cite{Edmonds}.  The final result
can be expressed in terms of Wigner $3j$- and $6j$--symbols
\cite{blumberg2}.  Here, we are merely interested in the dependence
of $|C(M,m,m_s)|^2$ on the magnetic quantum numbers for perpendicular
or $\sigma$--polarization of the laser beam. In this case,
\begin{widetext}
\begin{equation}
\label{eq:th16} C(M,m,m_s) = C \cdot \left(\begin{matrix}
     3/2 &  1   & 3/2\\
-(m+m_s) & \pm1 &   M\\
\end{matrix}\right)
\; \left(\begin{matrix}
2 & 1/2 & 3/2\\
m & m_s & -(m+m_s)\\
\end{matrix}\right)
\;,
\end{equation}
\end{widetext}
with $C$ being a constant weight parameter.  (The sign in the
$3j$--symbol should be chosen as to produce a non-vanishing result.)
The selection rules $m + m_s = M \pm 1$ and $|m + m_s| \leq \frac32$
enforced in (\ref{eq:th16}) leave twelve possible transitions grouped
into two sextets.  Their details are given in
Table~\ref{tab:transitions}. (If the restriction to
$\sigma$--polarization is lifted, additional transitions with $M =
m+m_s$ take place.) The total cross section $J_M(E)$ for an ion in
the initial magnetic state $|J,M\rangle$ then is obtained by
summation over all possible detachment channels:
\begin{eqnarray}
\label{eq:th17} J_{M}(\Omega) & = &
\sum_{m=-j}^{j}\sum_{m_s=-1/2}^{1/2} {\left|C(M,m,m_s)\right|}^2
\nonumber \\ & & \times J_{{\rm
eff}}\bigl[\Omega+\Delta\Omega(M,m,m_s)\bigr] \;.
\end{eqnarray}
As these transitions possess weight factors of comparable size,
details of the detachment spectrum invariably are lost in the
superposition of the individual currents $J_{{\rm eff}}[\Omega +
\Delta\Omega(M,m,m_s)]$ (\ref{eq:th11}).
\begin{table}
\begin{ruledtabular}
\begin{tabular}{ccccc}
$M$ & $m$ & $m_s$ & $|C_{\rm rel}(M,m,m_s)|^2$ & $\Delta\Omega(M,m,m_s)$ \\
\hline
$ \frac12$ & $ 1$ & $ \frac12$ &   3 & $ -1.8353 $\\
$-\frac12$ & $ 0$ & $ \frac12$ &   8 & $ -1.6682 $\\
$-\frac32$ & $-1$ & $ \frac12$ &   9 & $ -1.5012 $\\
$ \frac12$ & $ 2$ & $-\frac12$ &  12 & $ -1.3341 $\\
$-\frac12$ & $ 1$ & $-\frac12$ &  12 & $ -1.1671 $\\
$-\frac32$ & $ 0$ & $-\frac12$ &   6 & $ -1.0000 $\\
$ \frac32$ & $ 0$ & $ \frac12$ &   6 & $ +1.0000 $\\
$ \frac12$ & $-1$ & $ \frac12$ &  12 & $ +1.1671 $\\
$-\frac12$ & $-2$ & $ \frac12$ &  12 & $ +1.3341 $\\
$ \frac32$ & $ 1$ & $-\frac12$ &   9 & $ +1.5012 $\\
$ \frac12$ & $ 0$ & $-\frac12$ &   8 & $ +1.6682 $\\
$-\frac12$ & $-1$ & $-\frac12$ &   3 & $ +1.8353 $\\
\end{tabular}
\end{ruledtabular}
\caption{Details of the allowed transitions between the sublevels
$^2P_{3/2}$ in $S^-$ and $^3P_2$ in $S^0$. The fourth column denotes
the relative weight factor, the rightmost column the frequency shift in
units of $\omega_L$. \label{tab:transitions}}
\end{table}

A final correction that is rather negligible here but becomes
important for cooled ions concerns the temperature-dependent
occupation probabilities $P_M(T)$ of the initial ionic substates
$|J,M\rangle$ that, in thermal equilibrium, are given by the
Boltzmann distribution:
\begin{equation}
\label{eq:th18} P_M(T) = \rme^{- M G k_B T} \left[ \sum_{\mu=-J}^{J}
e^{-\mu G k_B T} \right]^{-1} \;.
\end{equation}
Finally, the experimentally measured quantity is the ratio
$R(\Omega,t_{\rm illu})$ of the number of surviving ions after a
period of illumination $t_{\rm illu}$ to the initial ion population
in the trap rather than the cross section itself.  Assuming a uniform
detachment rate, we find from (\ref{eq:th17}) after summing over all
initial ionic states:
\begin{equation}
\label{eq:th19} R(\Omega,t_{\rm{illu}}) = \sum_{M=-J}^{J} P_M(T)
\exp\bigl[-J_M(\Omega) t_{\text{illu}}\bigr] \;.
\end{equation}
The comparison of the results from this expression with the actual
experimental data is the subject of Section~\ref{sec:results}.

\section{EXPERIMENTAL TECHNIQUE}
\label{sec:experiment}

In the experiment we studied the effects of an external electric
field by observing the depletion of an $\rm S^{-}$ ion cloud stored
in a Penning ion trap \cite{blumberg1,blumberg2}.  To facilitate
comparison with the theory model, photodetachment data were acquired
as a function of photon energy both with and without the external
electric field. The ions were illuminated by a single-mode laser
tuned to the $\rm ^{2}P_{3/2} \rightarrow\, ^{3}P_{2}$ detachment
threshold.

The $\rm S^{-}$ ions are created in the Penning trap by dissociative
attachment to carbonyl sulfide (OCS) gas at a pressure of $7 \times
10^{-9}$ Torr, controlled by a variable leak.  The trap's magnetic
field is held fixed at 1.0~Tesla. The experiment is performed with a
trapped ensemble on the order of $10^{4}$ ions. A relative measure of
the number of trapped ions is made by resonantly driving the ions'
axial motion with a radio frequency voltage of 188~kHz applied to the
trap end caps.  The image current induced on the ring electrode is
detected at twice the driving frequency by a tuned antenna coil, and
the resulting voltage is then amplified in a heterodyne detection
scheme. The trap and its detection scheme provide a nearly noiseless
integrator of the trapped ion population, with noise introduced only
upon output of the signal \cite{blumberg1}.

\begin{figure}
\includegraphics{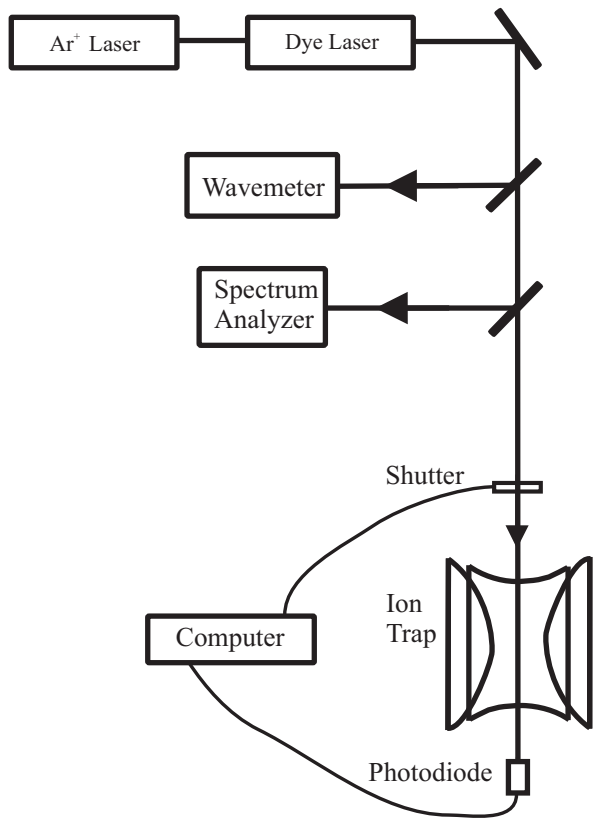}
\caption{\label{fig:apparatus} Schematic diagram of the major optical
apparatus and the ion trap. The ions undergo photodetachment with
light from a tunable, single-mode dye laser. A computer-controlled
mechanical shutter maintains a constant optical flux, measured by the
photodiode, from cycle to cycle of the data acquisition.}
\end{figure}

The ions are photodetached with light from a tunable laser.  The
laser system, shown in Fig.~\ref{fig:apparatus}, consists of a
standing-wave dye laser pumped by an $\rm Ar^{+}$ gas laser. The
dye laser uses rhodamine 6G dye to produce roughly 100~mW at
approximately 597~nm.  The continuously tunable output oscillates on
a single longitudinal mode.  The laser output is measured by a
travelling Michelson interferometer wavelength meter with a
resolution of 0.02~$\rm cm^{-1}$ \cite{hall76}.  The laser mode is
monitored by a Fabry-Perot spectrum analyzer with a 1500-MHz free
spectral range.  The laser beam is directed through the ion trap and
the transmitted flux is recorded by a photodiode.  A feedback signal
from the photodiode to a mechanical shutter maintains constant
integrated light intensity from cycle to cycle of the experiment.

The electric field is combined parallel to the trap's magnetic field
by adding a low radio frequency potential to the trap end caps.  A
typical frequency on the order of 1~MHz gives a field which is nearly
static on the time scale of the photodetached electrons'
36-picosecond cyclotron period.  Of course, this electric field is
not perfectly constant; thus a time average is calculated to quantify
the field strength.  The radio frequency is applied through a balun
transformer which insures opposite polarity on the end caps.  An
electrostatic analysis of the trap's hyperbolic electrodes shows that
the applied field is linear throughout the trapping volume.  The
electrostatic analysis also gives a measure of the field strength,
depending on the peak-to-peak applied radio frequency voltage.

One cycle of data acquisition consists of a background signal
measurement by the ion detection electronics, followed by an ion
creation period, an initial ion signal measurement, an optical
interaction period, and a final ion signal measurement.  The ion
creation period typically lasts 8~seconds and the optical interaction
typically lasts 500~ms. The ratio of the two ion signal measurements
yields the fraction of ions surviving the interaction period. To
account for trap losses by mechanisms other than detachment,
alternate data cycles measure the trap retention ratio in the absence
of laser light (shutter closed). Thus the ratio of the fraction of
ions surviving the interaction period with and without light yields a
fraction of ions surviving photodetachment, corrected to first order
for background losses. The entire cycle is repeated both with and
without the parallel electric field.

\section{RESULTS AND ANALYSIS}
\label{sec:results}

In the experiment, data were acquired to determine the ratio of $\rm
S^{-}$ ions surviving detachment for photon energies near the $\rm
^{2}P_{3/2} \rightarrow\, ^{3}P_{2}$ threshold. Observations were
made both with and without the external electric field parallel to
the magnetic field. The magnetic field was held at $B =1.0$~T, and
the peak applied radio frequency amplitude was $F_{\rm ext} =
14$~V/cm, for a r.~m.~s.\ average of 9.9~V/cm. The radio frequency
was 1.032~MHz, well above the trap eigenfrequency \cite{Footnote1}.
The single-mode laser beam was directed through the ion trap aligned
perpendicular to the fields and with $\sigma$~polarization, also
perpendicular to the fields.

\begin{figure}
\begin{center}
\includegraphics[width=0.9\columnwidth]{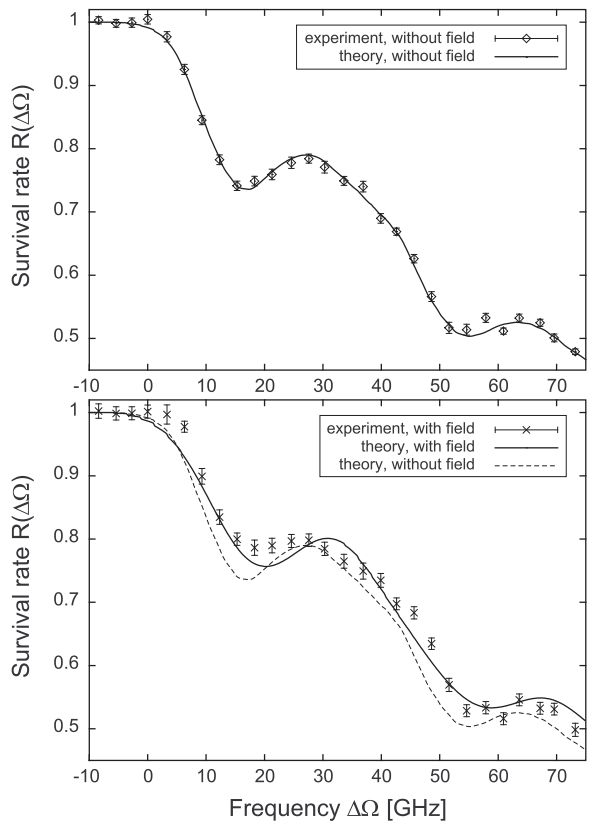}
\end{center}
\caption{Ratio of surviving ions $R(\Delta\Omega)$ (\ref{eq:th19}) in
photodetachment from S$^-$ in an external magnetic field as a
function of the laser detuning $\Delta\Omega$.  Top frame: without
electric field, bottom frame: with electric field. (Parameters used:
$B=1$~T, ion mass $M=32$~amu, $T=2100$~K, $F_{\rm ext} = 14$~V/cm
peak amplitude.)} \label{fig:Brate}
\end{figure}

In Fig.~\ref{fig:Brate} we show a comparison of the experimental and
theoretical results [Eq.~(\ref{eq:th19})] both with and without the
external electric field.  We plot the percentage of ions surviving
detachment as a function of the excess photon energy above the
electron affinity, expressed as a laser frequency detuning
$\Delta\Omega = \Omega - E_A/\hbar$. The top frame of
Fig.~\ref{fig:Brate} shows a comparison for the magnetic field-only
case (without the electric field). The bottom frame shows a
comparison for the combined parallel fields case.  To illustrate the
effect of the added electric field, the bottom frame also shows the
magnetic field-only theory curve.  In each case the initial threshold
and subsequent cyclotron structure are evident. As expected, due to
the large number of possible transitions
(Table~\ref{tab:transitions}) the degree of detail seen in
Fig.~\ref{fig:Baverage} is not easily recognized.  Fitting of the
theory curve to the data was optimized with an ion cloud temperature
of 2100 K and assuming an 8~$\deg$ rotation of the laser polarization
away from purely $\sigma$ polarization. Depending on the presence of
the additional electric field, the structure of both the theory and
experimental curves changes noticeably.  For both the experiment and
the theory,
addition of the electric field diminishes the depth of modulation in
the detachment ratio, particularly at the first dip above the initial
threshold. This dip corresponds directly to the first peak in the
relative detachment cross section shown in Fig.~\ref{fig:Baverage}.
We note in the bottom frame of Fig.~\ref{fig:Brate} that a larger
effect of the electric field is seen at the first dip for the
experimental data than for the theory curve, and this difference is
not well understood. However, the overall change in structure with
the electric field is to be expected from our theoretical analysis
(see Fig.~\ref{fig:Fparallel}), and is also qualitatively consistent
with our semi-classical view of the detachment process: the external
electric field pushes the electron away from the core as the electron
orbits in the magnetic field.  The result is a diminished overlap of
the electron wave function with itself, yielding a reduced cross
section. Furthermore, we observe that addition of the electric field
shifts the minimum of the first dip in the detachment ratio to
slightly higher energy, as predicted also by the theory.
Unfortunately, not all details of the experimental results obtained
for parallel fields could be reproduced by the theory.

Examination of the high precision achieved in experiments conducted
with ion beams (see, for example, Ref.~\cite{gibson93b,gibson2001})
might suggest that this work should be carried out in a beam
apparatus.  We considered this idea in depth, but found that
combining all the necessary field components, along with particle
detectors and optical access, was technically unfeasible.
Alternately, the Penning ion trap's size and geometric configuration
lends itself ideally to superposition of the parallel fields.
Furthermore, the trap's typical storage time allows for lengthy
optical interaction periods, and therefore greater precision in the
photon energy.

Of course, the trapped ions always experience the velocity-dependent
electric field perpendicular to the external magnetic field.  To
suppress its effects, efforts are currently underway to evaporatively
cool the trapped ion cloud. Future experiments at lower ion
temperatures are expected to yield greater contrast in the results
with the addition of the parallel electric field.  Also, detachment
experiments with $\pi$-polarized light (parallel to the fields) will
minimize the number of allowed Zeeman transitions and should give a
clearer picture of the underlying Green function.  Another possible
approach to diminish the number of Zeeman transitions is to quench
the detachment from the lowest three Zeeman sublevels of the ion
using a second laser.  This will require attention to the possibility
of collisional redistribution among the initial states, but again,
this effect should be diminished with an evaporatively cooled ion
cloud.

\section{CONCLUSIONS}
\label{sec:conclusions}

We have investigated, theoretically and experimentally,
near-threshold $s$--wave photodetachment from atomic negative ions in
parallel electric and magnetic fields.  Our fully quantum mechanical
theory based on the electron propagator allows us to calculate the
transition rate in combined fields for a thermal ion cloud in a
convenient manner, leaving only a single integral to be evaluated
numerically.  While showing agreement with previous theoretical work
for detachment from $\rm S^{-}$ in a purely magnetic field, the
theory predicts both a loss of contrast in the detachment rate and a
shift of the photocurrent maxima towards higher laser frequencies as
a consequence of the additional electric field. We present
experimental observations of photodetachment from $\rm S^{-}$ in
parallel fields that qualitatively agree with theoretical
expectations.  Future work will focus on similar experiments
conducted with evaporatively cooled ions, where effects of the
external electric field are expected to be more fully evident.

\begin{acknowledgments}
We wish to acknowledge helpful conversations with D.~J.~Larson.  This
work was sponsored by the Research Corporation, Davidson
College and the Deutsche Forschungsgemeinschaft (project number Kl
315/6-1). C.~B. would like to thank the Alexander von
Humboldt foundation and the Killam trust for financial support.
\end{acknowledgments}

\bibliography{praprl}

\begin{thebibliography}{46}
\bibitem[{Foo()}]{Footnote1}
\bibinfo{note}{A simple calculation shows that the effective external field
  $F_{\rm eff}$ acting on the trapped ions is given by $F_{\rm eff} = F_{\rm
  ext}\omega_{\rm ext}^2/(\omega_{\rm ext}^2 - \omega_{\rm trap}^2)$.}
\end{thebibliography}

\end{document}